\documentclass[11pt,a4paper]{article}
\usepackage{jheppub}
\usepackage{amsmath}
\usepackage[most]{tcolorbox}
\usepackage{dsfont}
\usepackage{ulem}
\usepackage{natbib}
\usepackage{xcolor}
\usepackage[hang,flushmargin]{footmisc}
\usepackage{tikz-cd}
\usepackage{enumitem}
\usepackage{amsfonts,amssymb, amscd,amsmath,latexsym,amsbsy,bm,pstricks}
\usepackage{stmaryrd}
\usepackage{todonotes}
\usepackage{stackrel}
\usepackage{amsthm}

\usepackage{graphicx}
\newcommand{\ti}{\textup i}

\title{\boldmath New Beta Integral from Supersymmetric Gauge Theory on Projective Space}

\author[a,c]{Ilmar Gahramanov}
\author[a,b]{Tuğba Hırlı}
\author[a,b]{R. Semih Kanber}
\author[d]{Hjalmar Rosengren}

\affiliation[a]{Department of Physics, Bogazici University, 34342 Bebek, Istanbul, Turkey}
\affiliation[b]{Department of Mathematics, Bogazici University, 34342 Bebek, Istanbul, Turkey}
\affiliation[c]{Center for Mathematics and its Apllications, Khazar University,  Mehseti St. 41, AZ1096, Baku, Azerbaijan}
\affiliation[d]{Department of Mathematical Sciences, Chalmers University of Technology and the University of Gothenburg, SE-412 96 G\"oteborg, Sweden}
\emailAdd{ilmar.gahramanov@bogazici.edu.tr}
\emailAdd{tugba3hirli@gmail.com}
\emailAdd{rcpsmh55@gmail.com}
\emailAdd{hjalmar@chalmers.se }

\abstract{We derive a new beta-type basic hypergeometric integral identity from the equality of supersymmetric partition functions on $\mathbb{RP}^{2}\times\mathbb{S}^{1}$. Unlike previously known identities obtained from lens-space partition functions, this integral does not appear to arise as a degeneration of the lens elliptic beta integral. Our result enriches the collection of basic hypergeometric beta integrals arising from supersymmetric dualities and has applications to supersymmetric gauge theories, integrable models, and the theory of special functions.}

\begin{document}

	\maketitle
	%\flushbottom

\section{Introduction}

In recent years, substantial progress has been made in the study of exactly computable observables in supersymmetric gauge theories formulated in diverse spacetime dimensions. This progress has been driven primarily by the method of supersymmetric localization, which allows for the exact evaluation of a wide class of nonperturbative quantities with rigorous analytical control. 

A particularly fruitful consequence of supersymmetric localization is the ability to test and refine proposed dualities between supersymmetric gauge theories. If two theories are conjectured to be dual, their partition functions on a given manifold must agree. Enforcing this equality often leads to highly nontrivial identities between special functions, many of which are highly intricate and difficult to discover or verify by purely mathematical means, see, for instance, \cite{Spiridonov:2009za,Spiridonov:2011hf,Gahramanov:2015tta,Gahramanov:2016wxi,Rosengren:2017ujl}. Thus, supersymmetric dualities provide not only physical insights but also a powerful source of new results in the theory of hypergeometric functions.

In this work, we derive a new beta-type basic hypergeometric integral identity\footnote{For a comprehensive survey of the one-variable basic hypergeometric beta integrals, see \cite{gasper2011basic,rahman1994pearson}.} that arises from the equality of partition functions in a pair of dual $\mathcal{N}=2$ supersymmetric gauge theories defined on the manifold $\mathbb{RP}^{2} \times \mathbb{S}^{1}$. The structure of this identity reflects the topological features of the background manifold as well as the matter content and gauge symmetry of the theories involved:
\begin{equation} \label{mainintegral}
\begin{aligned}
    & \frac{(q^{2};q^{2})_{\infty}}{(q;q^{2})_{\infty}}\, 
     \oint \frac{dz}{2\pi i\, z}\;
    \sum_{m=0}^{1}
    \left(
        q^{m}
        \prod_{i=1}^{3}
        \frac{
            \left( z^{-1}q^{m+1} a_i^{-1}; q^{2}\right)_{\infty}
            \left( z q^{m+1} b_i^{-1} ; q^{2}\right)_{\infty}
        }{
            \left( z q^{m} a_i ; q^{2}\right)_{\infty}
            \left( z^{-1} q^{m} b_i ; q^{2}\right)_{\infty}
        }
    \right),
    \\
    &\hspace{4cm}
    =
   \prod_{i=1}^{3} \prod_{j=1}^{3}
      \frac{
        \left( q a_i^{-1} b_j^{-1} ; q^{2} \right)_{\infty}
      }{
        \left( a_i b_j; q^{2} \right)_{\infty}
      },
\end{aligned}
\end{equation}
under the balancing condition $\prod_{i=1}^{3} a_i b_i = q$

From the physical perspective, we examine the $\mathbb{RP}^2 \times \mathbb{S}^1$ partition functions of three-dimensional $\mathcal{N}=2$ supersymmetric dual theories, originally introduced in \cite{Kashaev:2012cz}. Supersymmetric theories on such non-orientable manifolds are interesting for several reasons. They often exhibit novel topological features \cite{Kim:2013ola, LeFloch:2017lbt, Tanaka:2015igt}, modified parity structures, and unconventional localization behaviors that do not arise on orientable backgrounds. It appears that the index of theories on $\mathbb{RP}^{2} \times S^{1}$ does not admit the usual factorization \cite{Pasquetti:2011fj,Beem:2012mb,Hwang:2012jh}, since topologically $\mathbb{RP}^{2} \times S^{1}$
cannot be decomposed into two copies of $D \times S^{1}$ along their boundaries. In other words, the standard cutting-and-gluing construction used for $S^{2} \times S^{1}$, $S_b^3$ and $S^3/\mathbb{Z}_r$ fails on projective space, and the index does not factorize into a pair of holomorphic blocks in the conventional manner.

Furthermore, unlike the cases of $S^{2}\times S^{1}$ and lens spaces, no dimensional reduction from $S^{3}/\mathbb{Z}_{r}\times S^{1}$ to $\mathbb{RP}^{2}\times S^{1}$ is currently known. Consequently, the basic hypergeometric beta integral derived here does not appear to arise as a degeneration of the lens elliptic beta integral \cite{Kels:2015bda,Kels:2017toi,Gahramanov:2016ilb}.  In this context, the solution to the star-triangle equation presented in Sec.~3 does not appear to follow from the master solution of \cite{Kels:2015bda}.

%%%%%%%%%%%%%%%%%%%%%%%%%%%%%%%%%%%%%%%%%%%%%%%%%%%%%%%%5
\section{Duality for 3d $U(1)$ $\mathcal N = 2$ SQED}
%%%%%%%%%%%%%%%%%%%%%%%%%%%%%%%%%%%%%%%%%%%%%%%%%%%%%%%

We consider the supersymmetric index of a three-dimensional 
${\mathcal N}=2$ superconformal field theory defined on 
$\mathbb{RP}^{2} \times \mathbb{S}^{1}$ 
\cite{Tanaka:2014oda, Tanaka:2015pwa, Mori:2015urc}. 
It is given by
\begin{equation}
	I(q, \{a_j\}, \{n_j\}) 
	= \sum_{m_i} \frac{1}{|W_m|} 
	\oint \prod_{i=1}^{\mathrm{rank}\, G} 
	\frac{dz_i}{2\pi i\, z_i}\;
	Z_{\mathrm{gauge}}(z_i, m_i; q)\,
	Z_{\mathrm{chiral}}(z_i, m_i; a_j, n_j; q)\; .
\end{equation}
Here $Z_{\mathrm{gauge}}$ and $Z_{\mathrm{chiral}}$ denote the one-loop  contributions of the vector and chiral multiplets, respectively,  $|W_m|$ is the order of the Weyl group corresponding to a given monopole sector, the sum runs over all allowed monopole fluxes $m_i$, and the contour integral is taken over the gauge fugacities $z_i$.  This expression is completely fixed once the group-theoretical data of the theory, i.e. its global and gauge symmetries together with the representations of the matter fields are specified.

We consider the following three-dimensional ${\mathcal N}=2$ supersymmetric
duality\footnote{This supersymmetric duality was studied in a series of works 
\cite{Kashaev:2012cz,Gahramanov:2013rda,Gahramanov:2014ona,Gahramanov:2016wxi,Bozkurt:2020gyy}.}:

\medskip

\textbf{Theory A.}  
Supersymmetric electrodynamics with gauge group $U(1)$ and six chiral multiplets: three transforming in the fundamental representation and three transforming in the anti-fundamental representation of the gauge group.

\smallskip

The supersymmetric index (partition function) of Theory A takes the form
\begin{equation}
\label{eq:IA}
    \mathcal{I}_{A} 
    =  \frac{(q^{2};q^{2})_{\infty}}{(q;q^{2})_{\infty}}\, 
    q^{1/8} \oint \frac{dz}{2\pi i\, z}\;
    \sum_{m=0}^{1}
    \left(
        q^{-1/2+m}
        \prod_{i=1}^{3}
        \frac{
            \left( z^{-1}q^{m+1} a_i^{-1}; q^{2}\right)_{\infty}
            \left( z q^{m+1} b_i^{-1} ; q^{2}\right)_{\infty}
        }{
            \left( z q^{m} a_i ; q^{2}\right)_{\infty}
            \left( z^{-1} q^{m} b_i ; q^{2}\right)_{\infty}
        }
    \right),
\end{equation}
with fugacities obeying the balancing condition
\begin{equation} \label{balancingcond}
    \prod_{i=1}^{3} a_i b_i = q.
\end{equation}

\medskip

\textbf{Theory B.}  
A Wess-Zumino type model with no gauge degrees of freedom, containing nine gauge-invariant ``mesons'' transforming in the fundamental representation of the flavor symmetry group.

\smallskip

Its supersymmetric index is
\begin{equation}
\label{eq:IB}
    \mathcal{I}_{B}
    = q^{-3/8}
      \prod_{i=1}^{3} \prod_{j=1}^{3}
      \frac{
        \left( q a_i^{-1} b_j^{-1} ; q^{2} \right)_{\infty}
      }{
        \left( a_i b_j; q^{2} \right)_{\infty}
      },
\end{equation}
again subject to the same balancing condition.

The duality suggests that the supersymmetric index of these two theories is equivalent $\mathcal{I}_A= \mathcal{I}_B$, which corresponds to \eqref{mainintegral}.

%%%%%%%%%%%%%%%%%%%%%%%%%%%%%%%%%%%%%%%%%%%
\section{Star-Triangle relation}
%%%%%%%%%%%%%%%%%%%%%%%%%%%%%%%%%%%%%%%%%%%

The star-triangle relation is a fundamental equation in the theory of two-dimensional integrable lattice spin models in statistical mechanics. It provides the local condition that ensures the commutativity of transfer matrices and, consequently, the integrability of the model.

An interesting relationship, known as the gauge/YBE correspondence, has been identified in recent years; see the review articles \cite{Gahramanov:2017ysd,Yamazaki:2018xbx,Gahramanov:2022qge,Mullahasanoglu:2025ont} and references therein. It establishes a remarkable connection between solutions of the Yang-Baxter equation and partition functions of supersymmetric gauge theories. This correspondence provides a powerful framework for constructing new integrable models using field-theoretic techniques. In this setting, the equality of partition functions for the dual theories considered here naturally leads to a solution of the star-triangle equation.

In order to introduce spectral parameters\footnote{In the language of lattice models, these correspond to rapidity lines.}, we redefine the fugacities 
$a_i$ and $b_i$ as
\begin{equation}
    a_i = x_i^{-1} \alpha_i^{-1},
    \qquad 
    b_i = x_i\alpha_i^{-1} .
\end{equation}
With this parametrization, the identity takes the form
\begin{equation}
\begin{aligned}
    &\frac{(q^{2};q^{2})_{\infty}}{(q;q^{2})_{\infty}}\,
    q^{1/8}\oint \frac{dz}{2\pi i\, z}\;
    \sum_{m=0}^{1}
    \left(
        q^{-1/2+m}
        \prod_{i=1}^{3}
        \frac{
            \left(z^{-1} q^{m+1} x_i \alpha_i  ; q^{2} \right)_{\infty}
            \left( z\, q^{m+1} x_i^{-1}\alpha_i ; q^{2} \right)_{\infty}
        }{
            \left(z\, q^{m} x_i^{-1} \alpha_i^{-1} ; q^{2} \right)_{\infty}
            \left( z^{-1}  q^{m} x_i\,  \alpha_i^{-1} ; q^{2} \right)_{\infty}
        }
    \right)
    \\
    &\hspace{4cm}
    =\;
    q^{-3/8}
    \prod_{i=1}^{3} \prod_{j=1}^{3}
    \frac{
        \left( q\, x_i x_j^{-1}\alpha_i \alpha_j ; q^{2} \right)_{\infty}
    }{
        \left(x_j x_i^{-1} \alpha_i^{-1}  \alpha_j^{-1} ; q^{2} \right)_{\infty}
    }.
\end{aligned}
\end{equation}

We introduce the Boltzmann weight
\begin{equation}
    W_{\alpha}(x,z,q,m)
    =
    \frac{
        \left(z^{-1} q^{m+1} x \, \alpha \, ; q^{2} \right)_{\infty}
        \left(z\, q^{m+1} x^{-1} \alpha  ; q^{2} \right)_{\infty}
    }{
        \left( z q^{m} x^{-1} \alpha^{-1} ; q^{2} \right)_{\infty}
        \left( z^{-1}q^{m}x\, \alpha^{-1} ; q^{2} \right)_{\infty}
    },
\end{equation}
together with the spin-independent normalization factor
\begin{equation}
    R(\alpha_{1},\alpha_{2},\alpha_{3})
    =
    \prod_{i=1}^{3}
    \frac{(q\, \alpha_{i}^{2}; q^{2})_{\infty}}
         {(\alpha_{i}^{-2}; q^{2})_{\infty}}.
\end{equation}

In terms of these quantities, the identity takes the form of the star--triangle relation
\begin{multline}
    \frac{(q^{2};q^{2})_{\infty}}{(q;q^{2})_{\infty}}
    \oint \frac{dz}{2 \pi i\, z}\;
    \sum_{m=0}^{1}
        q^{m}\,
        W_{\alpha_{1}}(x_{1},z,q,m)\,
        W_{\alpha_{2}}(x_{2},z,q,m)\,
        W_{\alpha_{3}}(x_{3},z,q,m)
    \\
    =\;
    R(\alpha_{1},\alpha_{2},\alpha_{3})\,
    W_{\alpha_{1}\alpha_{2}}(x_{1},x_{2},q,0)\,
    W_{\alpha_{3}\alpha_{2}}(x_{3},x_{2},q,0)\,
    W_{\alpha_{1}\alpha_{3}}(x_{1},x_{3},q,0).
\end{multline}
As a result, we have constructed a new solution of the star-triangle equation. This solution defines a corresponding two-dimensional lattice model of statistical mechanics in which spins reside on the lattice sites and take both discrete (restricted to $0$ and $1$) and continuous values, interacting only with their nearest neighbors. Starting from this star-triangle solution, one can systematically build a solution of the Yang-Baxter equation; for instance, by following the steps outlined in \cite{Gahramanov:2015cva}. Likewise, one can obtain a solution of the star-star and star-square relation by adapting the constructions in 
\cite{Catak:2021coz,Mullahasanoglu:2023nes,Catak:2025hoz}.

\section{Pentagon relation}

The integral identity in (\ref{mainintegral}) can be rewritten in the form of a pentagon relation. Pentagon identities play an important roles in a wide range of mathematical and physical contexts. In particular, they encode the elementary  $2$-$3$ Pachner move, which represents the fundamental local transformation relating different triangulations of a $3$-manifold. In recent years, several nontrivial integral pentagon relations have been derived directly from three-dimensional $\mathcal{N}=2$ supersymmetric 
dualities; see, for instance, \cite{Dimofte:2011ju, Gahramanov:2016wxi,Benvenuti:2016wet}. 
%\footnote{There exist many interesting pentagon identities arising in mathematical physics, see, e.g., \cite{}.}

Let us define the function $\mathcal{B}$
\begin{equation}
    \mathcal{B}(a,b,c,m;q^2) = \frac{(q^{m+1} a^{-1} ;q^2)_\infty (q^{m+1} b^{-1} ;q^2)_\infty (q^{m+1} c^{-1};q^2)_\infty }{(q^m a;q^2)_\infty (q^m b;q^2)_\infty (q^m c;q^2)_\infty } .
\end{equation}
Then the duality on the level of superconformal indices can be written as follows
\begin{multline}
    \frac{(q^2;q^2)_\infty}{(q;q^2)_\infty}\oint \frac{dz}{2\pi iz} \sum_{m=0}^1 q^{m} \mathcal{B}(za_1 ,za_2,za_3,m;q^2)\mathcal{B}(z^{-1} b_1,z^{-1}b_2,z^{-1}b_3,m;q^2) \\ = \mathcal{B}(a_1 b_1 ,a_1 b_2,a_1 b_3,0;q^2) \mathcal{B}(a_2 b_1 ,a_2 b_2,a_2 b_3,0;q^2) \mathcal{B}(a_3 b_1 ,a_3 b_2,a_3 b_3,0;q^2).
\end{multline}
Similar pentagon identities have appeared in
\cite{Gahramanov:2013rda,Gahramanov:2014ona,Gahramanov:2016wxi,Bozkurt:2020gyy}.
It would be interesting to construct the corresponding Bailey pairs for the pentagon identity derived here, in analogy with the constructions in
\cite{Bozkurt:2020gyy,Gahramanov:2022jxz}, and to investigate their potential applications to the construction of knot invariants, along the lines of
\cite{Gahramanov:2025qrs}.

\section{Conclusions}

Basic hypergeometric integral identities of this type play an increasingly important role in mathematical physics. They arise naturally in the study of supersymmetric gauge theories, integrable lattice models, and quantum groups, with important applications in topological quantum field theory, knot theory, and the topology of three-manifolds. Since the identity derived in this work originates from supersymmetric gauge theories on the non-orientable manifold $\mathbb{RP}^{2}\times S^{1}$, it may provide a new link between these areas and serve as a starting point for further developments in the theory of basic hypergeometric functions, integrable systems, and quantum topology.

\section*{Acknowledgements}
Ilmar Gahramanov gratefully acknowledges Ali Pazarcı, Ege Çoban, and Mustafa Mullahasanoğlu for their valuable contributions during the initial stages of this work. Ilmar Gahramanov is supported by the Istanbul Integrability and Stringy Topics Initiative (\href{https://istringy.org}{istringy.org}).

\appendix
\section*{Appendix}
\addcontentsline{toc}{section}{Appendix}

\setcounter{section}{0}

\section{Proof of Duality}
In this part of the paper, we aim to prove the given duality $\mathcal{I}_A = \mathcal{I}_B$. We will prove the equality by establishing a more general one and showing that it holds in the basis of the vector space they live in. This allows us to conclude that, due to linearity, the equation will hold in general cases as long as the necessary conditions are satisfied. 

To proceed, we begin by recalling a known equality due to Gasper 
\cite{gasper:1989}, see also
\cite[Eq.\ (4.11.3)]{gasper2011basic}. By choosing appropriate values for the parameters in this equality, we obtain
\begin{multline}\label{gci}
\oint\frac{(zqb_1^{-1},qz^{-1}a_1^{-1};q)_\infty\theta(zqa_1^{-1} b_2b_3;q)}{(za_2,za_3,b_2z^{-1},b_3z^{-1};q)_\infty}\frac{dz}{2\pi\ti z}\\
=\frac{(qa_1^{-1}b_1^{-1};q)_\infty\prod_{i=2}^3(qa_i^{-1}b_1^{-1},qa_1^{-1}b_i^{-1};q)_\infty}{(q;q)_\infty\prod_{i,j=2}^3(a_ib_j;q)_\infty},
\end{multline}
where $\theta(a;q) = (a,q a^{-1} ;q)_{\infty} $, as long as we have the condition \eqref{balancingcond}. The contour of the integral is a deformed unit circle such that the poles of type $1/(za_2,za_3;q)_\infty$ lie outside of the contour, whereas the rest $1/(b_2/z,b_3/z;q)_\infty$ lie inside of the contour. The choice of the contour is also valid in the superconformal index for Theory A, so that it separates out the poles converging to infinity.

Building upon the previous result, we claim that this equality above can be generalized into 
\begin{multline}\label{gcig}\oint\frac{dz}{2\pi\ti z}\frac{f(z)}{(za_1,za_2,za_3,b_1z^{-1},b_2z^{-1},b_3z^{-1};q)_\infty}\\ 
=\frac{a_2f(a_2^{-1})\prod_{i=1}^3\theta(a_1b_i;q)-a_1 f(a_1^{-1})\prod_{i=1}^3\theta(a_2b_i;q)}{(q;q)_\infty\, a_2\theta(a_1a_2^{-1};q)\prod_{i,j=1}^3(a_ib_j;q)_\infty},
\end{multline}
where $f$ is any meromorphic function on $z\neq 0$ that satisfies 
\begin{equation}\label{quasi-per}f(qz)=-\frac{b_1b_2b_3}{q^2z^3}\,f(z).  \end{equation}
An alternative expression for the same integrals was given in \cite{Rosengren2016Rahman}.

By classical results on theta functions (see Corollary 1.3.5 of \cite{Rosengren:2016elliptic}), the space of admissible functions $f$ is three-dimensional and spanned by 
$$\theta(\mu_1z,\mu_2z,\mu_3z;q),$$
where $\mu_j$ are generic parameters subject to $\mu_1\mu_2\mu_3=q^2/b_1b_2b_3$. Therefore, since both sides of \eqref{gcig} are linear in \(f\), it is enough to verify the identity on a basis of this space.
In the special case $\mu=(q/b_1,a_1,q/a_1 b_2b_3)$, \eqref{gcig} reduces to \eqref{gci}. By symmetry, it holds in the six specializations $\mu=(q/b_j,a_k,qb_j/a_kb_1b_2b_3)$
where $j=1,2,3$ and $k=1,2$. These specializations contain three linearly independent functions, so they form a basis. Since the equality holds in this basis, it holds for every admissible function satisfying \eqref{quasi-per}.

We now take
$$f(z)=\prod_{i=1}^3\theta(zqa_i,qb_iz^{-1};q^2)+q\prod_{i=1}^3\theta(za_i,b_iz^{-1};q^2).$$

It is easy to verify that \eqref{quasi-per} is satisfied. The corresponding left-hand side of \eqref{gcig} can be written as
\begin{equation}\label{lhs}\oint\frac{dz}{2\pi\ti z}\sum_{m=0}^1 q^m\prod_{i=1}^3\frac{(q^{m+1}z^{-1}a_i^{-1},zq^{m+1}b_i^{-1};q^2)_\infty}{(zq^ma_i,q^mb_iz^{-1};q^2)_\infty}. \end{equation} 
On the right-hand side, the numerator is
\begin{multline*}
a_2\prod_{i=1}^3\theta(qa_ia_2^{-1},qa_2b_i;q^2)\theta(a_1b_i;q)-a_1\prod_{i=1}^3\theta(qa_ia_1^{-1},qa_1b_i;q^2)\theta(a_2b_i;q)\\
=\theta(q,qa_1a_2^{-1};q^2)\prod_{i=1}^3\theta(qa_1b_i,qa_2b_i;q^2)\\
\times\left\{a_2\theta(qa_3a_2^{-1};q^2)\prod_{i=1}^3\theta(a_1b_i;q^2)-a_1\theta(qa_3a_1^{-1};q^2)\prod_{i=1}^3\theta(a_2b_i;q^2)\right\},
\end{multline*}
where we used $\theta(a;q)=\theta(a,aq;q^2)$. 
To simplify this equation, we can use the following property related to the theta function, which is called the three-term identity \cite{Rosengren:2016elliptic}
\begin{equation} \label{threetermiden}
    \theta(bx^{\pm}, ac^{\pm};p)+\frac{a}{c}\theta(cx^{\pm},ba^{\pm};p)= \theta(ax^{\pm} + bc^{\pm};p),
\end{equation}
 where $\theta(ax^{\pm};q)=\theta(ax;q)\theta(a/x;q).
$
Then the numerator of the right-hand side simplifies into
$$a_2\theta(a_1a_2^{-1};q^2)\prod_{i=1}^3\theta(qa_3b_i;q^2). $$
Inserting this in \eqref{gcig}, it follows that \eqref{lhs} equals
$$\frac{\theta(q;q^2)}{(q;q)_\infty}\prod_{i,j=1}^3\frac{\theta(qa_i^{-1}b_j^{-1};q^2)_\infty}{(a_ib_j;q^2)_\infty}.$$
This proves the identity $\mathcal I_\mathcal A=\mathcal I_\mathcal B$.

	% BIBLIOGRAPHY
	% use BIBTEX if you want
	\bibliographystyle{JHEP}
	\bibliography{references}

	%\end{thebibliography}
\end{document}